\documentclass[a4paper,11pt]{article}

\usepackage{epsfig,multicol,amsmath}
\usepackage[T1]{fontenc}
\usepackage{jheppub}
\usepackage{amsmath}     
\usepackage{epsfig,epsf}
\usepackage{graphicx}
\usepackage{rotating}
\usepackage{verbatim}

\def\beq{\begin{equation}}
\def\eeq{\end{equation}}
\def\bea{\begin{eqnarray}}
\def\eea{\end{eqnarray}}

\def\eq#1{{Eq.~(\ref{#1})}}
\def\fig#1{{Fig.~\ref{#1}}}
\newcommand{\bas}{\bar{\alpha}_S}
\newcommand{\as}{\alpha_S}

\newcommand{\Lb}{\left(}
\newcommand{\Rb}{\right)}
\parskip=\itemsep               

\newcommand{\nn}{\nonumber}

\newcommand{\h}{\frac{1}{2}}

\newcommand{\pom}{I\!\!P}

\def\pom{{I\!\!P}}

\title{ CGC/saturation approach for soft interactions at high energy: 
inclusive production}
\author[a]{ E. ~Gotsman,}
\author[a,b]{ E.~ Levin}
\author[a]{  and U.~ Maor}

\affiliation[a]{Department of Particle Physics, School of Physics and Astronomy,
Raymond and Beverly Sackler
 Faculty of Exact Science, Tel Aviv University, Tel Aviv, 69978, Israel}
\affiliation[b]{Departemento de F\'isica, Universidad T\'ecnica Federico Santa Mar\'ia, and Centro Cient\'ifico-\\
Tecnol\'ogico de Valpara\'iso, Avda. Espana 1680, Casilla 110-V, Valpara\'iso, Chile}
\emailAdd{gotsman@post.tau.ac.il}
\emailAdd{leving@post.tau.ac.il, eugeny.levin@usm.cl}
\emailAdd{maor@post.tau.ac.il}

\abstract{ In this letter we demonstrate that our dipole model  is 
successful in describing
 the inclusive production within the same framework as  diffractive 
physics. We
 believe that this achievement stems from the fact that our approach 
incorporates the positive features
  of the Reggeon approach and CGC/saturation effective theory,
 for high energy QCD.}

\keywords{}
\begin{document}
\maketitle

\pagestyle{empty}

\pagestyle{plain}

\setcounter{page}{1}


\section{Introduction}

The LHC data on inclusive production \cite{ALICE,CMS,ATLAS} call for 
 a theoretical understanding of these processes within the framework of 
QCD. At first sight it would appear, that this process is a typical soft 
process, which occurs at
 long distances, where one should use the methods of non-perturbative 
QCD.
 Since such methods are only in an embryonic stage,  soft processes at 
high
 energy remain in  the arena of high energy phenomenology, based on the 
concept  of a soft Pomeron. Adopting this approach, inclusive 
production can be calculated using the technique of
 Mueller diagrams \cite{MUDI}. It has been demonstrated  that 
 soft Pomeron based models, provide a reasonable decription of the
 data\cite{GLMINCL1,GLMINCL2}. The advantage of our
 approach, is the feasibility of describing  inclusive production on 
the
 same footing as  diffractive production, and elastic scattering.

On the other hand, 
 in the CGC/saturation approach for 
iclusive production\cite{GLR,MUQI,MV,B,MUCD,K,JIMWLK}, one has a 
different senario.
 In this
 approach the inclusive production occurs in two stages. The first stage 
is
 the production of a mini-jet with the typical transverse momentum $Q_s$,
  where $Q_s$ is the saturation scale, which is much larger than the soft 
scale.
 This process is under full theoretical control. The second stage is the
 decay of the mini-jet  into hadrons,
which has to be treated phenomenologically,  using data from the hard
 processes. Such approach leads to a good description of the experimental
 data on inclusive production,  both for hadron-hadron, hadron-nucleus and
 nucleus-nucleus collisions, and observation of regularities in the 
data,
 such as geometric scaling  \cite{KLN,KLNLHC,LERE,MCLPR,PRA}. 
 The shortcoming of this approach is the fact that it is detached from
  diffractive physics.
  
  It should be mentioned, that the recently published measurements of the pseudorapidity 
distributions of charged particles in proton-proton collisions at an 
energy of 8 TeV, provide an additional challenge for model 
builders, which has not yet been successfully answered \cite{CMS}.

In this letter, we continue (see Refs.\cite{GLMNIM1,GLMNIM2} to construct 
a
 model for high energy soft interactions, which incorporates the 
advantages
 of both approaches. This model is based on the Colour Glass
 Condensate(CGC)/saturation  effective theory (see  Ref.\cite{KOLEB}
 for the review), and on the perturbative BFKL Pomeron\cite{BFKL}.
 We assume that 
the unknown mechanism for the confinement of quarks and gluons in QCD, is 
not 
 important, and its influence can be reduced to the determination of 
several
 parameters
related to the CGC/saturation approach, which depend on long distance 
physics. 

The main attributes of the model have been discussed in
 Refs.\cite{GLMNIM1,GLMNIM2},  in this paper we will 
 only include information that we  require for the discussion
 of inclusive production.


\section{Main formulae}

\begin{figure}[ht]
    \centering
  \leavevmode
      \includegraphics[width=13cm,height=5.5cm]{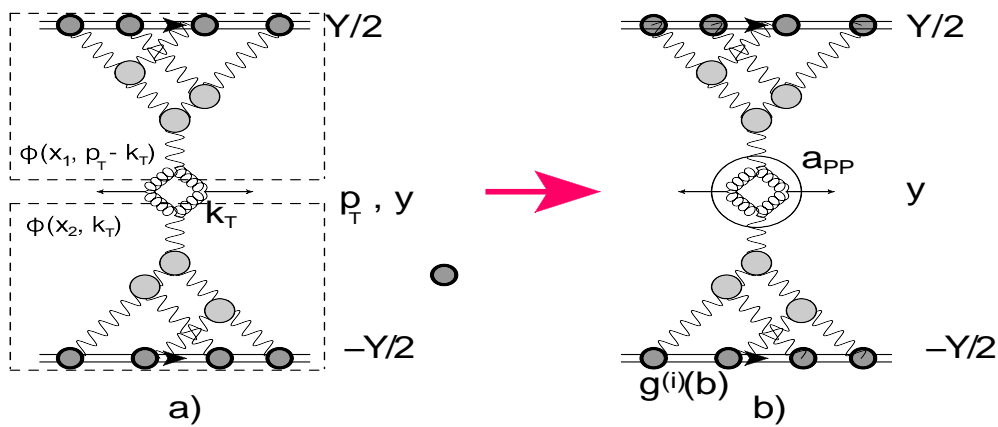}  
      \caption{The graphic representation of \protect\eq{CGCF}
 (see \fig{inclgen}-a).For the sake of simplicity all other indices
   in $\phi\Lb x_1,p_T - k_T\Rb$ and $\phi\Lb x_2,k_T\Rb$ are omitted.
 The wavy lines denote the BFKL Pomerons, while the helical lines 
illustrate
 the gluons. In \fig{inclgen}-b  the Mueller diagram for the inclusive
 production is shown. }
\label{inclgen}
   \end{figure}

First, we discuss the initial stage of  hadron production in the
 framework of CGC/saturation approach. For mini-jet production, we
 use the  $k_T$ factorization formula, that has been proven in  Ref.
 \cite{KTINC} ( see aslso Refs. \cite{BRINC,JMKINC,CMINC,KLINC,LPINC,KLPINC}
 where  this proof has been verified).

\beq \label{CGCF}
\frac{d \sigma}{d y \,d^2 p_T}\,\,=\,\,\frac{2 \pi \bas}{p^2_T}\int
 d^2 k_{T}\,\,\phi^{h_1}_G\Lb x_1;\vec{k}_T\Rb\,\phi^{h_2}_G\Lb x_2\vec{p}_T
 -\vec{k}_T\Rb
\eeq
where $\phi^{h_i}_G$ denotes the probability to find a gluon that
 carries the fraction $x_i$  of energy with $k_\perp$ transverse 
momentum,
 and $ \bas \,= \,\as N_c/\pi$,  with the number of colours equal to
 $N_c$. $\h Y + y \,=\,\ln(1/x_1)$ and $ \h Y - y =\ln(1/x_2)$.
 $\phi^{h_i}_G$ is the solution of the Balitsky-Kovchegov(BK)
 \cite{B,K} non-linear evolution equation, and can be viewed as
 the sum of `fan' diagrams of the BFKL Pomeron interactions,  shown
 in \fig{inclgen}. 

In our model we  use  the simple formula which is a good 
approximation to
 the numerical solution of the BK equation, see
 Ref.\cite{LEPP}: viz.

 \beq \label{SOLNU}
N^{BK}\Lb G_\pom\Lb z\Rb\Rb \,\,=\,\,a\,\Lb 1
 - \exp\Lb -  G_\pom\Lb z\Rb\Rb\Rb\,\,+\,\,\Lb 1 - a\Rb
\frac{ G_\pom\Lb z\Rb}{1\,+\, G_\pom\Lb z\Rb},
\eeq 
 with $a = 0.65$ and $G_\pom\Lb z \Rb\,\,=\,\,\phi_0\Lb r^2 Q^2_s\Lb y, b \Rb\Rb^{1 - \gamma_{cr}}$ 
 where we have used two inputs:
 $r = R$ and $Q^2_s\,=\,\Lb 1/(m^2R^2)\Rb\,S\Lb m\,b\Rb\,\exp\Lb \lambda y\Rb
 $.
 For the values of $1 - \gamma_{cr}$ and $\lambda $, we have estimates in 
the leading
 order of perturbative QCD: $1 - \gamma_{cr}\,=\,0.63$ and
 $\lambda \,=\,4.88\bas$.
  The value of $\lambda$ is a fitting parameter, which
 effectively includes the higher order QCD corrections.  
 In this paper we use the value $\lambda = 0.38$ which we found in 
Ref.\cite{GLMNIM2}.
The parameter $m$ and the function $S\Lb m \,b\Rb$  originate from
  non-perturbative QCD contributions,  and   are taken in the following
 form:
 \beq \label{SDI}
S \Lb m \,b\Rb \,\,=\,\,\frac{m^2}{\pi^2} \,e^{- m\,b}~~~~\mbox{where}~~~~
\int d^2 b \,S\Lb b\Rb\,=\,1
\eeq
   $\phi_0$ can be calculated using  the initial 
conditions, of the  BFKL equation.

  However, we do not know these conditions, and so we consider $\phi_0$ as 
 an
 additional phenomenological parameter.   The values of these parameters
 are taken from Ref.\cite{GLMNIM2}: $m = 5.25 GeV$ and $\phi_0 = 0.0019$.
 All these parameters describe the CGC /saturation structure of the BFKL
 Pomerons and their interactions. We introduce phenomenological parameters
 to describe the structure of the hadron. We choose the two channel model
 for such a structure and describe the vertex of the BFKL Pomeron 
interaction with
 a hadron state $i$ in the following form:
   \beq \label{G}
g_i\Lb m_i, b\Rb\,\,=\,\,g_i \,S_\pom \Lb m_i, b\Rb ~~~~~~\mbox{where}~~~
~S_\pom \Lb m_i, b\Rb \,\,=\,\, \,\frac{1}{4\pi}\,m^3_i\,b\,K_1\Lb m_i\,b \Rb
\eeq
  The parameters that we use  in this paper, have been extracted
 from fitting  the elastic and diffractive data in
 Ref.\cite{GLMNIM2}, and their values are:  
   \beq \label{PARA}
   g^{(1)}\,=\,110.2 \,GeV^{-1};~~~m_1\,=\,0.92\,GeV;~~ 
  g^{(2)}\,=\,11.2 \,GeV^{-1};~~~m_2\,=\,1.9\,GeV;   
   \eeq
   
   Finally, \eq{CGCF} can be re-written as a Mueller diagram
 of \fig{inclgen}-b, and the inclusive cross section is given by
   
   \bea \label{INCF}
   \frac{d \sigma}{d y}\,\,&=&\,\,\int d^2 p_T\,\frac{d
 \sigma}{d y \,d^2 p_T}\,\,=\,\,a_{\pom \pom}\,\ln\Lb W/W_0\Rb\Bigg\{
 \alpha^4  \,In^{(1)}\Lb \h Y + y\Rb \, In^{(1)}\Lb \h Y - y\Rb  \,\nn\\
   &+&\,\alpha^2\beta^2 \Big(In^{(1)}\Lb \h Y + y\Rb \, In^{(2)}\Lb
 \h Y - y\Rb \,+\,  In^{(2)}\Lb \h Y + y\Rb \, In^{(1)}\Lb \h Y -
 y\Rb\Big)\,\nn\\
   &+&\,\beta^4 \,In^{(2)}\Lb \h Y + y\Rb \, In^{(2)}\Lb \h Y - y\Rb 
 \Bigg\}
   \eea
    where $\alpha$ and $\beta$ describe the structure of the diffractive
 scattering in the two channel model, where the observed physical 
hadronic and diffractive states are written in the form 
\beq \label{MF1}
\psi_h\,=\,\alpha\,\Psi_1+\beta\,\Psi_2\,;\,\,\,\,\,\,\,\,\,\,
\psi_D\,=\,-\beta\,\Psi_1+\alpha \,\Psi_2;~~~~~~~~~
\mbox{where}~~~~~~~ \alpha^2+\beta^2\,=\,1;
\eeq 
$In^{(i)}$ is given by
\beq \label{IN}
In^{(i)}\Lb y\Rb\,=\,\int d^2 b \,\,N^{BK}\Lb g^{(i)}\,S\Lb m_i,
 b\Rb \,\tilde{G}_\pom\Lb y\Rb\Rb
\eeq
where  $\tilde{G}_\pom\Lb y\Rb\,=\,\,\phi_0 \exp\Lb \lambda\Lb 1
 - \gamma_{cr}\Rb y\Rb$ and $N^{BK}$ is defined in \eq{SOLNU}.
 Regarding the factor in front of \eq{INCF}  i.e. $ \ln\Lb
 W/W_0\Rb$, where $W = \sqrt{s}$ is the energy of collision in c.m.f.,
 and $W_0$ is the  value of energy from which we can start our approach.
  One can see that \eq{CGCF}
is divergent in the region of small $p_T \,<\,Q_s$. Indeed,
 in this region $\phi$'s in \eq{CGCF} do not depend on $p_T$,
 since $k_T \approx\,Q_s \,>\,p_T$, and the integration over $p_T$
 leads to $\ln\Lb Q^2_s/m^2_{soft}\Rb$, where $m_{soft}$ is the
 non-perturbative scale that includes the confinement of quarks
 and gluons ($m_{soft} \sim \Lambda_{QCD})$.

To convert the rapidity distribution (which we calculate theoretically),
 to pseudorapidity ($\eta$) one, we need to know the mass of min-jet 
($m$). The simple
 formulae are  well known (see Ref. \cite{KLN} for example):
\beq \label{PO2}
y\Lb \eta, p_T\Rb \,\,=
\,\,\frac{1}{2} \ln\left\{\frac{\sqrt{\frac{ m^2_{jet}+ 
 p^2_T}{ p_T^2} \,+\,\sinh^2\eta}\,\,+\,\,\sin\eta}{\sqrt{\frac{ m^2_{jet}
  +  p_T^2}{p^2_\perp}\,+\,\sinh^2\eta}\,\,-\,\,\sinh \eta}\right\} 
  \eeq
with the Jacobian
\beq \label{PO3}
h\Lb \eta,p_T\Rb \,\,= \frac{\cosh\eta}{\sqrt{\frac{m^2_{jet}  + 
 p^2_T}{ p^2_T}\  \,+\,\sinh^2\eta}}
\eeq 
 The mass of mini jet is given by $m^2_{jet} = 2 m_{soft}  p_T$
 (see  Ref. \cite{KLN}).  Since the typical transverse momentum
 is equal to the saturation scale, we have
 \beq \label{PO4}
 \frac{m^2_{jet}}{p^2_T}\,=\,\frac{2 \,m_{soft}}{Q_s\Lb W \Rb}\,\,=\,\,r^2_0
 \,\Lb \frac{W}{W_0}\Rb^{ - \h \lambda}
 \eeq
 where $r^2_0$  and $a_{\pom \pom}$  are phenomenological parameters 
that are determined 
from
 the experimental data.

 Finally,
 \beq \label{ETA} 
 \frac{d \sigma}{d \eta}\,\,=\,\,h\Lb \eta, Q_s\Rb \frac{d \sigma}{ d y}\Lb y\Lb \eta, Q_s\Rb \Rb 
 \eeq
 
  \section{Comparison with the experimental data}
\begin{figure}
\begin{tabular}{c c}
 \includegraphics[width=80mm]{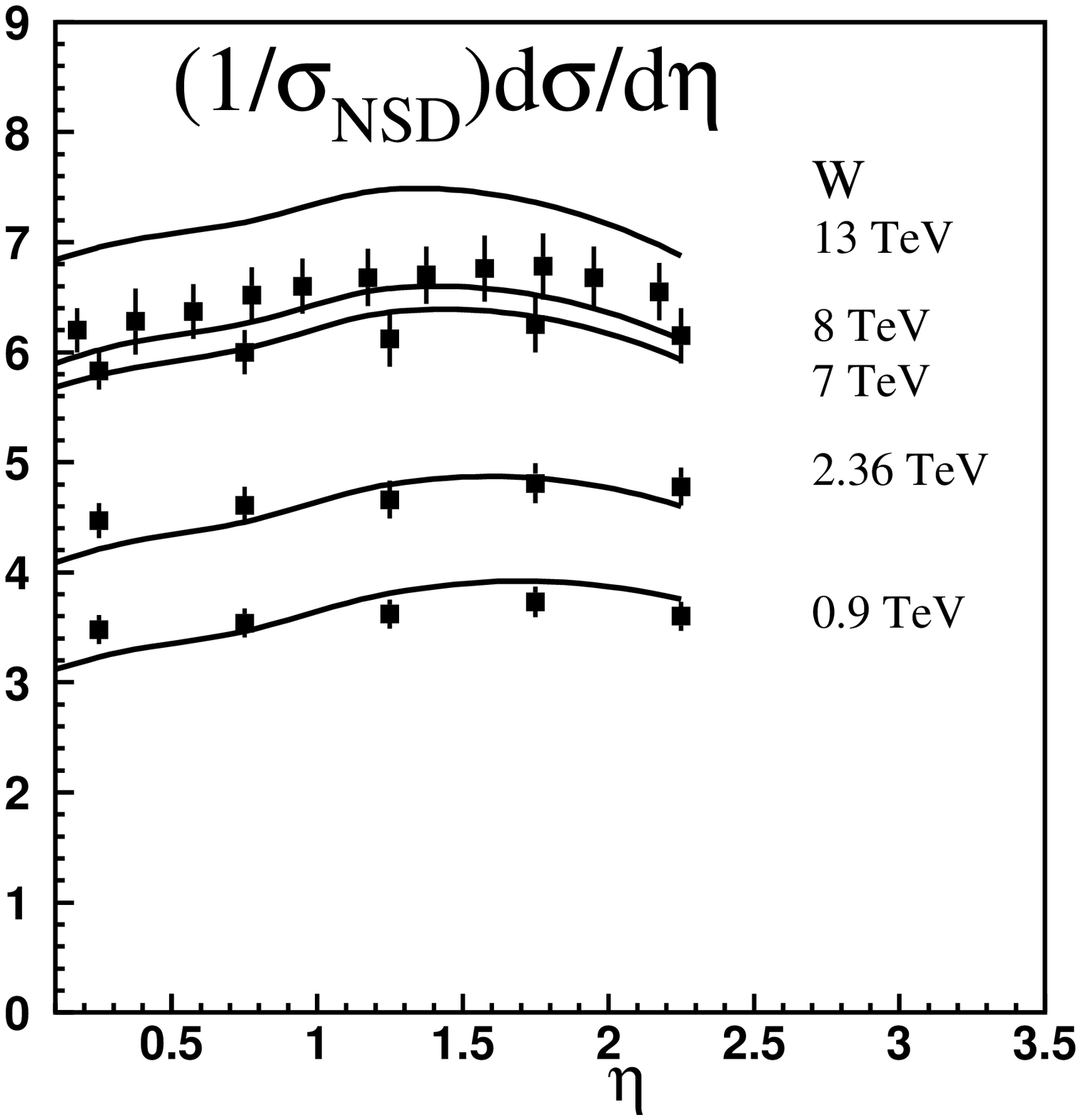}&\includegraphics[width=80mm]{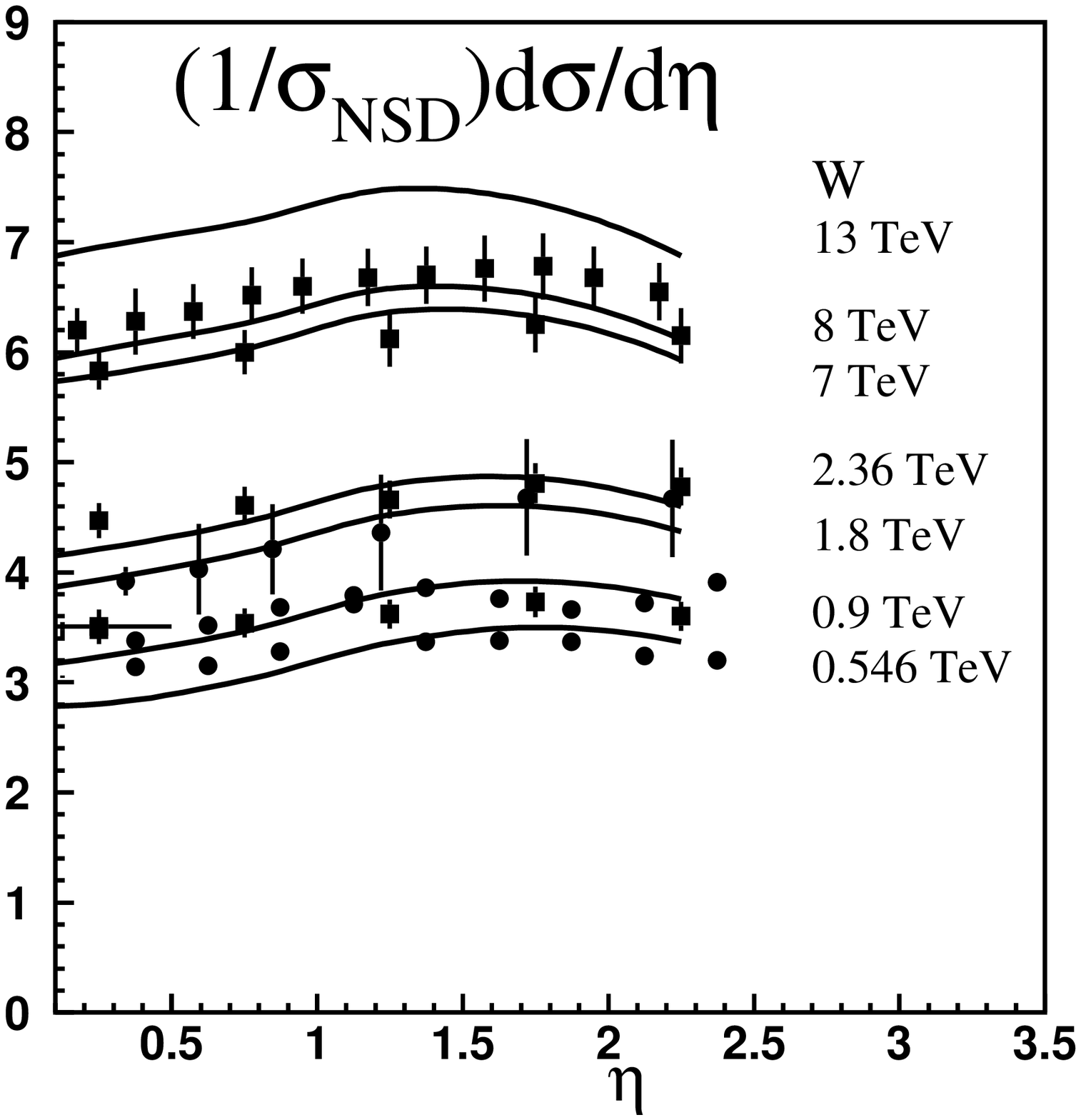}\\ 
\fig{incl}-a & \fig{incl}-b\\
\end{tabular}
\caption{The single inclusive density ( $(1/\sigma_{NSD})\,d \sigma/d \eta$) versus energy.  
The data were taken from Refs.\cite{ALICE,CMS,ATLAS} 
and from Ref.\cite{PDG}. description of the multi-particle production in hard processes in 
The description of  the CMS data is plotted in \fig{incl}-a, 
while \fig{incl}-b presents the comparison with  all inclusive spectra 
with $W \geq 0.546 \,TeV$. }
\label{incl}
\end{figure}

  In \fig{incl} we plot our predictions compared to the  experimental 
data. 
As we 
have mentioned all other parameters have been extracted from the 
diffractive and
 elastic data in Ref.\cite{GLMNIM2}. The only free parameters
 are $a_{\pom\pom}$ and $r^2_0$.  The curves in \fig{incl} are calculated 
for $a_{\pom \pom} = 0.21$ and $r^2_0 \, = \,8 $.
  
  From  \eq{INCF}, we note that the inclusive cross section is 
sensitive 
to the contribution of the black component.
 As we discussed in Ref.\cite{GLMNIM2}, qualitatively,
  we have  in our two channel model two different components:
 one which is transparent, even at ultra high energy
 (e.g. at $W = 57\, TeV$)
 while the second component, starts  being black at rather low energy 
(say at $W=0.9\,TeV$). 
Hence, our good description of the experimental data,
 checks that the value of this component is consistent with the inclusive 
data.
  
\begin{figure}[ht]
     \centering
\begin{tabular}{c c}
    \includegraphics[width=78mm]{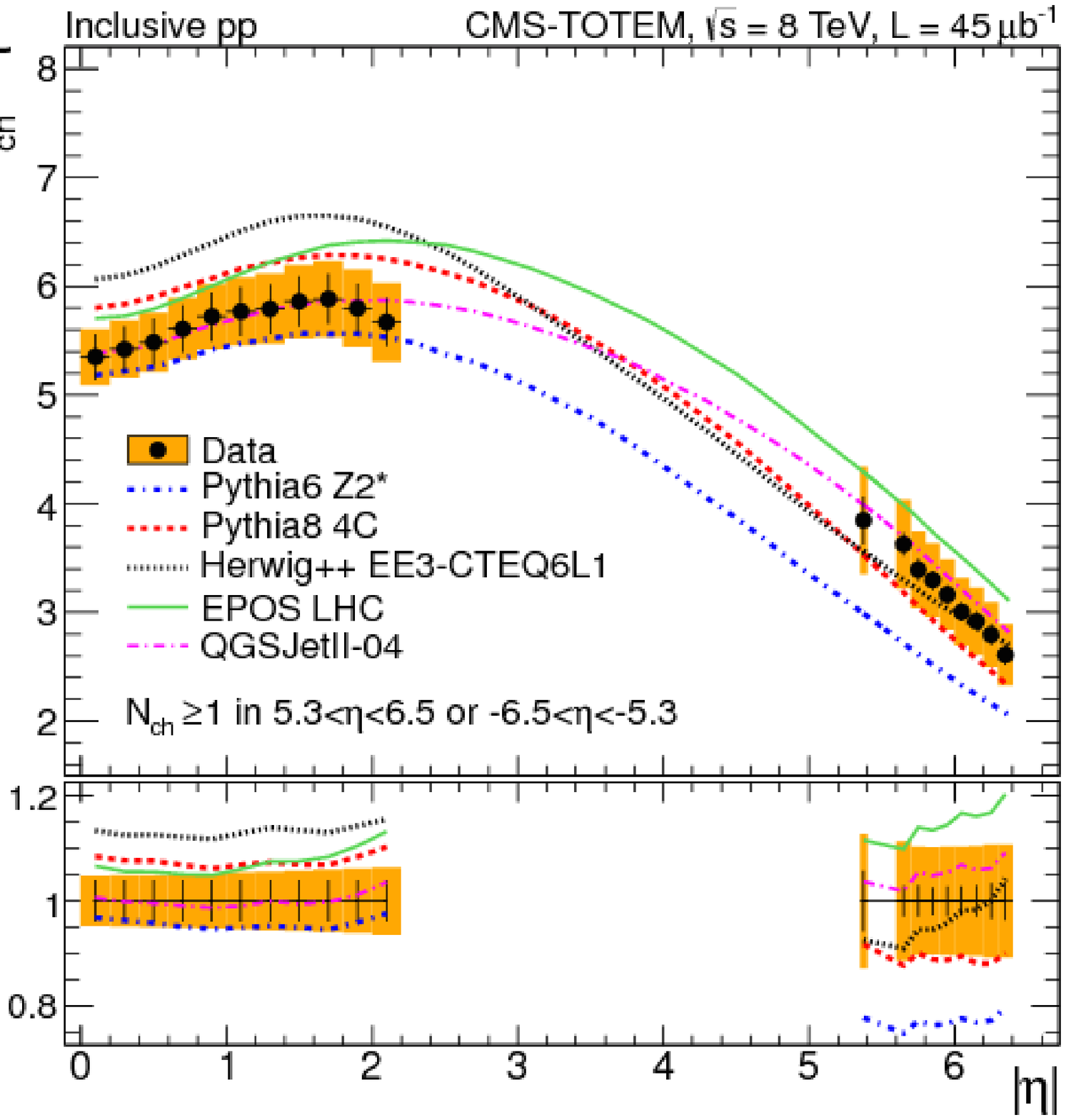} &  \includegraphics[width=80mm]{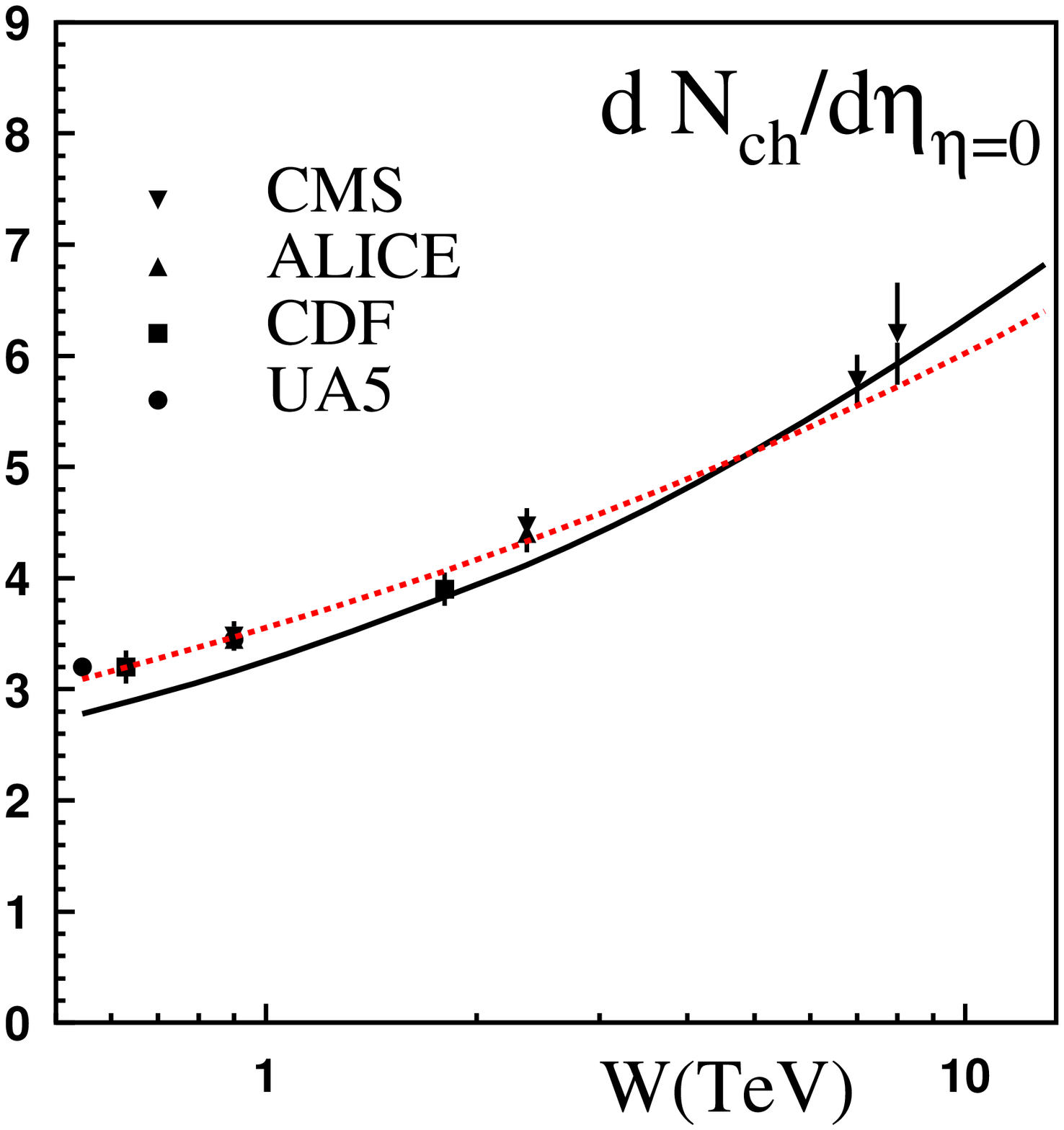}  \\
 \fig{incly0}-a &\fig{incly0}-b\\
    \end{tabular}
      \caption{The comparison of the inclusive production at $W = 8\,TeV$ with the Monte Carlo models is shown in \fig{incly0}-a. The picture is taken from Ref.\cite{CMS}. In \fig{incly0}-b it shown  $d N_{ch}/d \eta$ at $\eta = 0$ versus energy $W$. Our  estimates are shown by the solid line. The dotted line corresponds to fit: $0.725 \Lb W/W_0\Rb^{0.23} $ with $W_0 = 1\,GeV$. The data are taken from Refs.\cite{CMS,ALICE,CDF,UA5}. }
\label{incly0}
   \end{figure}

  One can see that our model describes the value of the inclusive 
 densities $\rho = (1/\sigma_{NSD})d \sigma/d \eta$ and their dependence on
 energy and rapidity, rather well. It should be stressed that the values 
for 
$\sigma_{NSD}$ were calculated in
  our model. Regarding the new data at W= 8 TeV (Fig. 2 and 3), the  
comparison  shows that the result
of our approach,
 is slightly below  the experimental central values, while the 
numerous
 Monte Carlo simulations  overshoot the data in the central region (see 
\fig{incly0}-a).
 
  Note, we have only dealt with data in the central region, since we
do not take into account parton correlations due to energy conservation.
These are important in the fragmentation region, but difficult to 
incorporate in our present framework. 
  
In \fig{incly0}-b it is shown the energy dependence  of  $d N_{ch}/d \eta$ at $\eta = 0$. In the  CGC/saturation approach $d N_{ch}/d \eta|_{\eta=0}
\,\,\propto\,W^\lambda$,  where $\lambda$ corresponds to the energy dependence of the saturation scale. In our model the energy dependence is more complicated and can be approximate as $W^{0.29}$. Note, that the power $0.29$ is much less than the value of $\lambda $ ($\lambda$ = 0.38). Recall, that in our model $\lambda$ characterizes the energy behavior of the saturation scale.

 Concluding this discussion we see that we reached a good description of the data but our estimates are below of the data at small values of rapidity $\eta$. We believe, that this is a reflection of our simplified relation between $y$ and $\eta$.

    \section{Conclusions}  
  In this letter we demonstrate that our model for the soft (long distance) interaction which is based on CGC/saturation approach, is able to describe  inclusive production. In other words, we give the example that our model can describe both the diffractive (elastic) physics at high energy and the typical production process which in the majority of the approaches are treated in different ways. We believe that  such a successful approach roots in the fact, that our procedure incorporates the advantages of high energy phenomenology based on the soft Pomeron interactions , and of the CGC/saturation effective theory that
includes the description of multi-particle production in perturbative QCD.

On the other hand, this letter is the natural next step in our search for a model based on QCD, that will be able to describe the typical  properties of high energy interaction and to include the diffractive production and multi-particle  generation process on the same footing. It is also a next step in our attempts to build this description without addressing the Monte Carlo simulation methods.
  
  \section*{Acknowledgements}
   We thank our colleagues at Tel Aviv university and UTFSM for
 encouraging discussions. Our special thanks go to Carlos Contreras,
 Alex Kovner and Misha Lublinsky for elucidating discussions on the
 subject of this paper.
   This research was supported by the BSF grant   2012124  and the
 Fondecyt (Chile) grant 1140842.



\begin{thebibliography}{99}
\bibitem{ALICE}
 K.~Aamodt {\it et al.}  [ALICE Collaboration],
  Eur.\ Phys.\ J.\ C {\bf 68} (2010) 89
  [arXiv:1004.3034 [hep-ex]];\,\,
ALICE Collaboration,
Eur.\ Phys.\ J.\  C {\bf 65} (2010) 111
[arXiv:0911.5430 [hep-ex]].
\bibitem{CMS}
S.~Chatrchyan {\it et al.}  [CMS and TOTEM Collaborations],
  Eur.\ Phys.\ J.\ C {\bf 74} (2014) 10,  3053
  [arXiv:1405.0722 [hep-ex]];\,\,
  V.~Khachatryan {\it et al.}  [CMS Collaboration],
  Phys.\ Rev.\ Lett.\  {\bf 105}, 022002 (2010;\,\,
  [arXiv:1005.3299 [hep-ex]].
   V.~Khachatryan {\it et al.}  [CMS Collaboration],
  JHEP {\bf 1002} (2010) 041
  [arXiv:1002.0621 [hep-ex]].
\bibitem{ATLAS}
ATLAS Collaboration,
arXiv:1003.3124 [hep-ex].
\bibitem{MUDI}
A. H. Mueller,
{\it Phys. Rev.} {\bf D2} (1970) 2963.

\bibitem{GLMINCL1}
 E.~Gotsman, E.~Levin and U.~Maor,
  Phys.\ Rev.\ D {\bf 81}, 051501 (2010)
  [arXiv:1001.5157 [hep-ph]].
\bibitem{GLMINCL2}
E.~Gotsman, E.~Levin and U.~Maor,
  Phys.\ Rev.\ D {\bf 84}, 051502 (2011)
  [arXiv:1103.4509 [hep-ph]].
%
\bibitem{GLR}
L. V. Gribov, E. M. Levin and M. G. Ryskin, 
Phys. Rep. {\bf 100} (1983) 1. 
\bibitem{MUQI}
A. H. Mueller and J. Qiu, 
Nucl. Phys. {\bf B268} (1986) 427.
\bibitem{MV}
L. McLerran and R. Venugopalan, 
Phys. Rev. {\bf D49} (1994) 2233, 3352; {\bf D50} (1994) 2225; 
{\bf D53} (1996) 458;\\ {\bf D59} (1999) 09400. 

\bibitem{B}
I.~Balitsky,
[arXiv:hep-ph/9509348];\,\,
{\it Phys.\ Rev.} {\bf D60}, 014020 (1999)
[arXiv:hep-ph/9812311]\,\,\,\,
\bibitem{MUCD}
A. H. Mueller,
Nucl. Phys. {\bf B415} (1994) 373; {\bf B437} (1995) 107.

\bibitem{K}
Y.~V.~Kovchegov,
{\it Phys.\ Rev.}  {\bf D60}, 034008  (1999),
[arXiv:hep-ph/9901281].
\bibitem{JIMWLK}
~J.~Jalilian-Marian, A.~Kovner, A.~Leonidov and H.~Weigert,
{\it  Phys.\ Rev.}\,  {\bf D59}, 014014 (1999),
[arXiv:hep-ph/9706377];\,\,  {\it Nucl.\ Phys.}\,{\bf B504}, 415
(1997),
[arXiv:hep-ph/9701284]; \,\,\,
J.~Jalilian-Marian, A.~Kovner and H.~Weigert,
  {\it Phys.\ Rev.}  {\bf D59}, 014015 (1999),
  [arXiv:hep-ph/9709432];\,\,\,
 A.~Kovner, J.~G.~Milhano and H.~Weigert,
 {\it  Phys.\ Rev.}  {\bf D62}, 114005 (2000),
  [arXiv:hep-ph/0004014]\,; \,\,\,
E.~Iancu, A.~Leonidov and L.~D.~McLerran,
{\it  Phys.\ Lett.}\,  {\bf B510}, 133 (2001);
[arXiv:hep-ph/0102009];\,\, {\it  Nucl.\ Phys.}\,  {\bf A692}, 583
(2001),
[arXiv:hep-ph/0011241];\,\,\,
E.~Ferreiro, E.~Iancu, A.~Leonidov and L.~McLerran,
 {\it  Nucl.\ Phys.}\  {\bf A703}, 489 (2002),
  [arXiv:hep-ph/0109115];\,\,\,
H.~Weigert,
{\it  Nucl.\ Phys.}  {\bf A703}, 823 (2002),
[arXiv:hep-ph/0004044].
%
\bibitem{KLN}
D.~Kharzeev, E.~Levin and M.~Nardi,
  Nucl.\ Phys.\  A {\bf 730} (2004) 448
  [Erratum-ibid.\  A {\bf 743} (2004) 329]
  [arXiv:hep-ph/0212316];\,\
  Phys.\ Rev.\  C {\bf 71} (2005) 054903
  [arXiv:hep-ph/0111315];\,\, D.~Kharzeev and E.~Levin,
  Phys.\ Lett.\  B {\bf 523} (2001) 79
  [arXiv:nucl-th/0108006];\,\,\, D.~Kharzeev and M.~Nardi,
  Phys.\ Lett.\  B {\bf 507} (2001) 121
  [arXiv:nucl-th/0012025].



\bibitem{KLNLHC}
  D.~Kharzeev, E.~Levin and M.~Nardi,
  Nucl.\ Phys.\  A {\bf 747} (2005) 609
  [arXiv:hep-ph/0408050].

\bibitem{LERE}
 E.~Levin and A.~H.~Rezaeian,
  AIP Conf.\ Proc.\  {\bf 1350} (2011) 243
  [arXiv:1011.3591 [hep-ph]];\,\,
  Phys.\ Rev.\ D {\bf 82} (2010) 014022
  [arXiv:1005.0631 [hep-ph]]; \,\, 
[arXiv:1102.2385 [hep-ph]];\,\,
{\it Phys. Rev} {\bf D82} (2010) 054003.
[arXiv:1007.2430 [hep-ph]].

\bibitem{MCLPR}
L. McLerran, M. Praszalowicz,
  Nucl.\ Phys.\ A {\bf 916} (2013) 210
  [arXiv:1306.2350 [hep-ph]];\,\,
{\it Acta Phys. Polon.} {\bf B42 } (2011) 99 
[arXiv:1011.3403 [hep-ph]];\,\,
{\it Acta Phys. Polon.} {\bf B41} (2010) 1917.
[arXiv:1006.4293 [hep-ph]].
\bibitem{PRA}
M. Praszalowicz,
  Phys.\ Lett.\ B {\bf 727} (2013) 461
  [arXiv:1308.5911 [hep-ph]];\,\,
  Acta Phys.\ Polon.\ Supp.\  {\bf 6} (2013) 3,  809
  [arXiv:1304.1867 [hep-ph]];\,\,
  Phys.\ Lett.\ B {\bf 704} (2011) 566
  [arXiv:1101.6012 [hep-ph]]l\,\,
  Phys.\ Rev.\ Lett.\  {\bf 106} (2011) 142002
  [arXiv:1101.0585 [hep-ph]].


\bibitem{GLMNIM1}
E.~Gotsman, E.~Levin and U.~Maor,
  Eur.\ Phys.\ J.\ C {\bf 75} (2015) 1,  18
  [arXiv:1408.3811 [hep-ph]].
\bibitem{GLMNIM2}
 E.~Gotsman, E.~Levin and U.~Maor,
  arXiv:1502.05202 [hep-ph].
\bibitem{KOLEB}
Yuri V Kovchegov and Eugene Levin, {\it `` Quantum Choromodynamics at High Energies"}, Cambridge Monographs on Particle Physics, Nuclear Physics and Cosmology, Cambridge University Press, 2012 .
\bibitem{BFKL}
 E. A. Kuraev, L. N. Lipatov, and F. S. Fadin, {\it  Sov. Phys.
JETP}
                {\bf 45}, 199 (1977); \,\,\,
Ya. Ya. Balitsky and L. N. Lipatov,
               {\it   Sov. J. Nucl. Phys.}\, {\bf 28}, 22 (1978).








  \bibitem{KTINC}
   Y.~V.~Kovchegov and K.~Tuchin,
  Phys.\ Rev.\  D {\bf 65} (2002) 074026
  [arXiv:hep-ph/0111362].
\bibitem{BRINC} 
M.~A.~Braun,
  Eur.\ Phys.\ J.\  C {\bf 48} (2006) 501
  [arXiv:hep-ph/0603060];\,\,\,
 Phys. Lett. B  {\bf 483} (2000), 105.
\bibitem{JMKINC}
   J.~Jalilian-Marian and Y.~V.~Kovchegov,
  Phys.\ Rev.\  D {\bf 70} (2004) 114017
  [Erratum-ibid.\  D {\bf 71} (2005) 079901]
  [arXiv:hep-ph/0405266].
\bibitem{CMINC}
   C.~Marquet,
  Nucl.\ Phys.\  B {\bf 705} (2005) 319
  [arXiv:hep-ph/0409023].
\bibitem{KLINC}
 A.~Kovner and M.~Lublinsky,
  JHEP {\bf 0611} (2006) 083
  [arXiv:hep-ph/0609227].
  \bibitem{LPINC}
   E.~Levin and A.~Prygarin,
  Phys.\ Rev.\  C {\bf 78} (2008) 065202
  [arXiv:0804.4747 [hep-ph]].
  \bibitem{KLPINC}
A.~Kormilitzin, E.~Levin and A.~Prygarin,
  Nucl.\ Phys.\  A {\bf 813} (2008) 1
  [arXiv:0807.3413 [hep-ph]].

\bibitem{LEPP}
E.~Levin,
  JHEP {\bf 1311} (2013) 039
  [arXiv:1308.5052 [hep-ph]].

\bibitem{PDG}
C. Amsler et al. (Particle Data Group), 
{\it Phys. Lett.} {\bf B667} (2008) 1. 
\bibitem{CDF}
 T.~Aaltonen {\it et al.}  [CDF Collaboration],
  Phys.\ Rev.\ D {\bf 79} (2009) 112005
   [Erratum-ibid.\ D {\bf 82} (2010) 119903]
  [arXiv:0904.1098 [hep-ex]];\,\,
 F.~Abe {\it et al.}  [CDF Collaboration],
  Phys.\ Rev.\ D {\bf 41} (1990) 2330.
\bibitem{UA5}
 G.~J.~Alner {\it et al.}  [UA5 Collaboration],
  Z.\ Phys.\ C {\bf 33} (1986) 1.
\end{thebibliography}
\end{document}